\documentclass{emulateapj}
\newcommand{\Mpc}{\mbox{ Mpc}}

\newcommand{\Mpcinv}{\mbox{ Mpc$^{-1}$}}

\newcommand{\secinv}{\mbox{ s$^{-1}$}}

\newcommand{\Msun}{\mbox{ M$_\odot$}}

\newcommand{\Lsun}{\mbox{ L$_\odot$}}

\newcommand{\hunits}{\mbox{ km s$^{-1}$ Mpc$^{-1}$}}

\newcommand{\lya}{Ly$\alpha$ }

\newcommand{\bq}{\begin{equation}}
\newcommand{\eq}{\end{equation}}
\newcommand{\bqa}{\begin{eqnarray}}
\newcommand{\eqa}{\end{eqnarray}}
\def\VEV#1{\left\langle #1\right\rangle} 

\begin{document}

\title{Spatial Correlations in the Helium-Ionizing Background}

\author{Steven R.  Furlanetto}

\affil{Department of Physics and Astronomy, University of California, Los Angeles, CA 90095, USA; email: sfurlane@astro.ucla.edu}

\begin{abstract}
After quasars ionize intergalactic \ion{He}{2} at $z \sim 3$, a large radiation field builds up above the \ion{He}{2} ionization edge.  Unlike the background responsible for \ion{H}{1} ionizations, this field should be highly variable, thanks to the scarcity of bright quasars and the relatively short attenuation lengths ($\sim 50 \Mpc$) of these high-energy photons.  Recent observations of the \ion{He}{2} and \ion{H}{1} \lya forests show that this background does indeed vary strongly, with substantial fluctuations on scales as small as $\sim 2 \Mpc$.  Here we show that such spatial fluctuation scales are naturally expected in any model in which the sources are as rare as bright quasars, so long as the attenuation length is relatively small.  The correlation length itself is comparable to the attenuation length ($\ga 10 \Mpc$) for the most plausible physical scenarios, but we find order-of-magnitude fluctuations on all scales smaller than $\sim 6 \Mpc$.  Moreover, aliasing along the one-dimensional skewers probed by the \ion{He}{2} and \ion{H}{1} \lya forests exaggerates these variations, so that order-of-magnitude fluctuations should be \emph{observed} on all scales smaller than $\sim 20 \Mpc$.  Complex radiative transfer is therefore not required to explain the observed fluctuations, at least at the level of current data.
\end{abstract}
  

\section{Introduction} \label{intro}

To the vast majority of baryonic matter in the Universe, the most important radiation field is the metagalactic ionizing background, and a great deal of effort has gone into understanding its properties (e.g., \citealt{rauch97, mcdonald01, tytler04, bolton05, fan06, becker07}).  Most work has focused on measuring the mean amplitude of this background, which is indeed the only interesting aspect if the background is spatially uniform.  This appears to be an excellent assumption for \ion{H}{1}-ionizing photons at $z \la 4$, where spatial fluctuations are probably only a few percent of the mean \citep{zuo92a, fardal93, meiksin04, croft04, bolton05, furl08-hefluc}, although it breaks down near and during the reionization epoch at $z \ga 6$ \citep{meiksin03, bolton07, mesinger08-ib, choudhury08, furl08-mfp}.  

Spatial fluctuations are much more important for photons above the ionization edge of \ion{He}{2}, which can be studied with the \ion{He}{2} \lya forest at $z \sim 3$ (e.g., \citealt{miralda93, jakobsen94}).  A particularly powerful approach is to compare the \ion{He}{2} and \ion{H}{1} \lya forests along a given line of sight.  The hardness ratio $\eta = N_{\rm HeII}/N_{\rm HI}$ parameterizes the strengths of each individual absorber; it depends on the ratio of the local ionization rates for hydrogen $\Gamma_{\rm HI}$ and helium $\Gamma_{\rm HeII}$ \citep{miralda93}, which are in turn integrated measures of the ionizing background.  Several recent studies suggest that the \ion{He}{2}-ionizing background fluctuates strongly at $z \sim 2.6$, with nearly order-of-magnitude spatial variations on scales spanning a few to a few tens of Mpc \citep{shull04, zheng04, fechner06, fechner07} and some evidence for hardening to higher redshifts \citep{heap00}.  

Large fluctuations are naturally expected for two reasons.  First, only quasars are able to produce photons with energies above the 54.4 eV ionization threshold of \ion{He}{2}, and these sources are relatively rare -- implying a strongly fluctuating background even after reionization \citep{fardal98, bolton06, furl08-hefluc}.  Direct evidence for these source-induced variations has been seen in the ``transverse proximity effect" of the hardness ratio through comparisons of the \ion{H}{1} and \ion{He}{2} \lya forests with surveys for nearby quasars \citep{jakobsen03, worseck06, worseck07}.  This is exaggerated by the strong attenuation from residual intergalactic \ion{He}{2}:  the mean free path of these photons (at $\ga 30 \Mpc$) is about an order of magnitude smaller than that of \ion{H}{1}-ionizing photons \citep{haardt96, faucher08-ionbkgd, furl08-helium}, so a typical point in the intergalactic medium (IGM) sees only a few sources.  Second, radiative transfer through the clumpy IGM can induce additional fluctuations \citep{maselli05, tittley07}.

Naively, one might expect random variations in the local quasar population to imprint a characteristic scale comparable to the attenuation length, attributing smaller scale variations to radiative transfer effects.   Here we will quantitatively examine the spatial scales of correlations induced in the \ion{He}{2} ionizing background by stochastic variations in quasar source counts.  We will show that the $1/r^2$ intensity profiles around the discrete sources create strong small-scale fluctuations and suggest that the variations observed in the data may not require complex radiative transfer.

In our numerical calculations, we assume a cosmology with $\Omega_m=0.26$, $\Omega_\Lambda=0.74$, $\Omega_b=0.044$, $H_0=100 h \hunits$ (with $h=0.74$), $n=0.95$, and $\sigma_8=0.8$, consistent with the most recent measurements \citep{dunkley08,komatsu08}.   Unless otherwise specified, we use comoving units for all distances.

\section{The Correlation Function of the Ionizing Intensity}
\label{corr}

We define the correlation function $\xi_J(r)$ of the amplitude $J$ of the ionizing background (for concreteness, at the ionization edge of \ion{He}{2}) via
\bq
\VEV{J({\bf r}_1) J({\bf r}_2)} = \VEV{J}^2 [1 + \xi_J(r)],
\label{eq:corrfcn}
\eq
where ${\bf r}_{1,2}$ label two points in the IGM, $r = | {\bf r}_1 - {\bf r}_2|$, $\VEV{..}$ denotes a spatial average, and we have assumed isotropy.  For Poisson distributed sources and a fixed photon attenuation length $r_0$, the probability distribution function of $J({\bf r})$, as well as the joint distribution function at two spatial points, can be derived either via Markoff's method \citep{zuo92a, zuo92b} or the method of characteristic functions \citep{meiksin03}; see \citet{furl08-hefluc} for an application to the \ion{He}{2}-ionizing background.  \citet{zuo92b} shows how to compute the correlation function from these distributions:
\bq
\xi_J(r) = {1 \over 3 \bar{N}_0} \, {\VEV{L^2} \over \VEV{L}^2} \, {r_0 \over r} \, I_J(r/r_0),
\label{eq:corrJ}
\eq
where $\bar{N}_0 = (4 \pi/3) n_Q r_0^3$ is the mean number of quasars inside an attenuation volume, $n_Q$ is the total number density of quasars, and 
\bq
I_J(u) = \int_0^\infty dx {x \over \sinh x} \exp \left( -u \, {1 + e^{-x} \over 1 - e^{-x}} \right).
\label{eq:IJ}
\eq
In equation~(\ref{eq:corrJ}), $L$ denotes the quasar luminosity, and its moments are computed by integrating over the source luminosity function $\Phi(L)$.  Here we have made several simplifying assumptions:  (i) quasars counts are Poisson-distributed around the mean value $\bar{N}_0$ (see \citealt{furl08-hefluc} for a discussion of deterministic clustering); (ii) sources are visible to infinite distance, attenuated by a factor $\tau=r/r_0$ (see \citealt{zuo92b} for a discussion of fluctuations in attenuation); and (iii) the effective volume is Euclidean (certainly reasonable for $r_0 \la 300 \Mpc$, as used here).  This corresponds to the post-reionization limit in \citet{furl08-hefluc}.

The function $I_J(u)$ is shown in \citet{zuo92b}.  It limits to $\pi^2/4$ as $u \rightarrow 0$, so $\xi_J \propto r^{-1}$ for $r \ll r_0$.  For $r \gg r_0$, it is exponentially suppressed.  Thus the shape of $\xi_J$ is determined entirely by our underlying assumptions (and in particular the $e^{-(r/r_0)}/r^2$ flux profile of randomly distributed sources); the luminosities and number densities of the sources only affect the amplitude of the correlation function (although real world complexities, and especially radiative transfer, may modify this simple expectation).

However, to determine the amplitude we need $\VEV{L^2}$ and $\VEV{L}$, and thus $\Phi(L)$.  This is relatively well-determined at the redshifts of interest ($2 \la z \la 3$), but unfortunately our results are extremely sensitive to it. The quasar luminosity function is typically parameterized as a double power law (e.g., \citealt{pei95, boyle00, hopkins07}),
\bq
\Phi(L) = {\Phi_\star/L_\star \over (L/L_\star)^{-\alpha} + (L/L_\star)^{-\beta}},
\label{eq:dplaw}
\eq 
where $\Phi_\star$ normalizes the density, $L_\star$ is the characteristic break luminosity, $\alpha \sim -1.5$, and $\beta \sim -3$; all of these constants are redshift-dependent.  At $L \gg L_\star$, $\Phi \propto L^{\beta}$; thus $\VEV{L^2}$ diverges if $\beta \ge -3$ and converges only slowly if $\beta \la -3$.  

A finite maximum quasar luminosity $L_{\rm max}$ somewhat alleviates this problem.  As a reasonable guess, we use the Eddington luminosity for a $10^{11} \Msun$ black hole (corresponding to a bolometric luminosity $3.3 \times 10^{15} \Lsun$), converted to the $B$-band using the correction of \citet{hopkins07}.  This is comparable to the brightest observed quasars, or $\sim 200 L_\star$ for the redshifts of interest.  Doubling $L_{\rm max}$ increases the amplitude of $\xi_J$ by $\sim 5\%$.\footnote{Note that $\bar{N}_0$ and the moments of $L$ are also sensitive to the minimum quasar luminosity, but this dependence nearly cancels from $\xi_J$.}

Unfortunately, the bright-end slope still introduces substantial uncertainty.  To gauge its impact, consider the recent \cite{hopkins07} luminosity function estimated from a wide range of observational samples.  The best-fit bolometric luminosity function has $\beta \approx -3.3$ at $z=2.5$.  On the other hand, if the fitting parameters are forced to vary smoothly with redshift, $\beta \approx -3.1$.  If a (non-linear) correction is then applied to reconstruct the $B$-band values, the luminosity function flattens farther to $\beta \approx -2.9$.  Finally, a naive correction from $B$-band luminosity to the helium ionization edge using the observed range of quasar UV spectral indices \citep{zheng98, telfer02, scott04} yields $\beta \approx -2.8$ \citep{furl08-hefluc}, although this procedure may overestimate the flattening if the $B$-band correction is correlated with the far-UV scatter.

For concreteness, we use the \citet{hopkins07} $B$-band function that varies smoothly with redshift as our fiducial model ($\beta \approx -2.9$) but show some results for other plausible values of $\beta$.  Note that we only require the total number counts $\bar{N}_0$ and the shape of the luminosity function, so we do not need to correct $\Phi(L)$ to the far-UV explicitly.  When varying $\beta$, we hold $n_Q$ constant.

\begin{figure}
\plotone{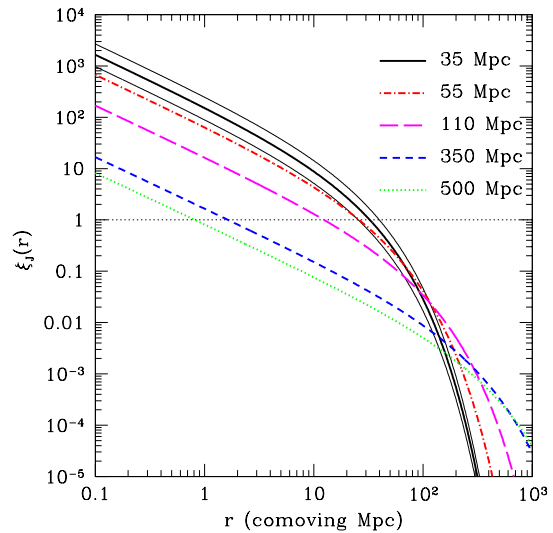}
\caption{Correlation function of the ionizing intensity, $\xi_J(r)$.  The thick curves use our fiducial quasar luminosity function at $z=2.5$ and take $r_0=35,\,55,\,110,\,350,$ and 500 Mpc, from top to bottom.  The thin solid curves show the plausible range of uncertainty from $\Phi(L)$.}
\label{fig:corr}
\end{figure}

Figure~\ref{fig:corr} shows the resulting correlation functions at $z=2.5$.  From top to bottom, the thick curves take $r_0=35,\,55,\,110,\,350,$ and 500 Mpc.  The first two are plausible values for \ion{He}{2}-ionizing photons (see \citealt{furl08-helium, furl08-hefluc}); the others span the range expected for \ion{H}{1}-ionizing photons \citep{madau99-qso, faucher08-ionbkgd}.  

With $\Phi(L)$ held fixed between these curves, the amplitude at $r \ll r_0$ is $\propto r_0/\bar{N}_0 \propto r_0^{-2}$.  The shapes, when normalized to $r_0$, are identical:  $\xi_J \sim r^{-1}$ at small separations followed by an exponential suppression at $r \sim r_0$.  Measuring this cutoff would therefore be an effective measure of the attenuation length.

However, $r_0$ is \emph{not} the typical scale at which points are highly correlated ($\xi_J \sim 1$).  That is set by the number density of sources and decreases with $r_0$.  We find the correlation lengths to be $(32,26,12,1.6,0.8) \Mpc$ for $r_0=(35,55,110,350,500) \Mpc$, respectively.  Only when $r_0 \sim 30 \Mpc$ is the correlation length larger than the attenuation length.  

Instead, the correlation length is closer to (though not exactly equal to) the typical radius at which a quasar dominates the local photoionization rate, the so-called ``proximity zone."  For \ion{He}{2}-ionizing photons, this is $R_{\rm prox} \sim 16 \Mpc$ at $z=2.5$ if $L_B=10^{12} \Lsun$ (near $L_\star$) and $\Gamma_{\rm HeII} \approx 5 \times 10^{-15} \secinv$ (a reasonable value for the ionizing background at $z \sim 2.5$; \citealt{sokasian02, bolton06}); $R_{\rm prox}$ is at least an order of magnitude smaller for \ion{H}{1}-ionizing photons.  Strong correlations are confined to the proximity zones around quasars; they are extremely important for the \ion{He}{2} photons simply because that background fluctuates so strongly \citep{bolton06, furl08-hefluc}.

The thin solid curves in Figure~\ref{fig:corr} assume $r_0=35 \Mpc$ but take different values for $\beta$ in $\Phi(L)$:  the upper curve has $\beta \approx -2.8$ (corresponding to a naive inclusion of the scatter in quasar UV spectral; \citealt{furl08-hefluc}), while the lower curve follows the shape of the $z=2.5$ (only) bolometric fit in \citet{hopkins07}, with $\beta \approx -3.3$.  The total uncertainty in the amplitude of $\xi_J$ therefore spans at least a factor of three, comparable to the differences from the plausible range of $r_0$.

Note that we have assumed here that quasars produce the entire ionizing background.  While most likely true for \ion{He}{2}-ionizing photons (though see the discussion of recombination radiation in \S \ref{caveat}), galaxies provide much (if not most) of the \ion{H}{1}-ionizing background (e.g., \citealt{faucher08-ionbkgd}).  Galaxies are so common that their fluctuations will be tiny \citep{furl08-hefluc}, and the additional background can be considered uniform.  In that case, $\xi_J$ is suppressed by a factor $(J_{\rm QSO}/J_{\rm tot})^2$, and the correlation length will decrease as well.

\section{The Power Spectrum of the Ionizing Intensity}
\label{powspec}

A cleaner tool for separating the scale-dependence of the correlations is the (three-dimensional) Fourier transform of the correlation function, the power spectrum $P_J(k)$:\footnote{Actually, a simpler way to compute the power spectrum is with the halo model \citep{cooray02}; our physical model of uncorrelated sources with constant attenuation is equivalent to ignoring the two-halo term and inserting a ``halo" profile $\propto e^{-(r/r_0)}/r^2$ \citep{scherrer91}.}
\bq
P_J(k) = 4 \pi \int_0^\infty dr \, \xi_J(r) {r \sin(kr) \over k}.
\eq  

\begin{figure}
\plotone{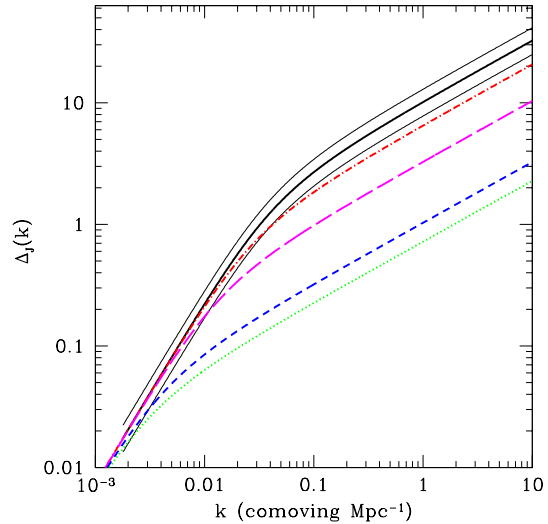}
\caption{Three-dimensional power spectrum of the ionizing intensity, normalized to dimensionless form $\Delta_J(k) = [(k^3/2 \pi^2) P_J(k)]^{1/2}$.  All thick curves assume $z=2.5$ and take $r_0=35,\,55,\,110,\,350,$ and 500 Mpc, from top to bottom (line styles are identical to Fig.~\ref{fig:corr}); the thin solid curves show the plausible range of uncertainty from $\Phi(L)$.}
\label{fig:pk}
\end{figure}

We show the dimensionless form of this quantity, $\Delta_J(k) = [(k^3/2 \pi^2) P_J(k)]^{1/2}$, in Figure~\ref{fig:pk}; this is roughly equal to the fractional rms scatter when the ionizing background is smoothed on a scale $\ell = 2\pi/k$.  The line styles are identical to those in Figure~\ref{fig:corr}, with $r_0$ increasing from top to bottom.

As one should expect, $\Delta_J$ has a break at $k_0 = 2 \pi/r_0$.  At $k>k_0$, $\Delta_J \propto k^{1/2}$, which follows from the $1/r$ dependence of $\xi_J$.  For $k < k_0$, $\Delta_J \propto k^{3/2}$, thanks to the exponential suppression in $\xi_J$.  

Unfortunately, measuring this power spectrum, and hence identifying this feature, requires three-dimensional data.  For studying the ionizing background, we are instead limited to one-dimensional skewers along the rare \lya forest lines of sight to UV-bright quasars.  Under these conditions, we can only measure the one-dimensional power spectrum,
\bq
P_{J, \rm 1D}(k_\parallel) = \int_{k_{\parallel}}^{k_{\rm max}} {dk \over 2 \pi} k P_J(k),
\label{eq:pk_1d}
\eq
where the integration over smaller physical scales accounts for the possibility of aliasing.

This immediately presents a problem:  at $k>k_0$, $P_J \propto k^{-2}$, so the integral is logarithmically divergent.  Clearly the observed power \emph{at every scale} will be dominated by the smallest scales for which the power is non-zero.  In reality, this limit will be set by Jeans smoothing and thermal broadening in the IGM, which together smear out power on smaller scales.  For concreteness, we set $k_{\rm max} = 10 \Mpc^{-1}$ in our calculations, comparable to the Jeans scale for fully-ionized mean density gas at $z=2.5$.

\begin{figure}
\plotone{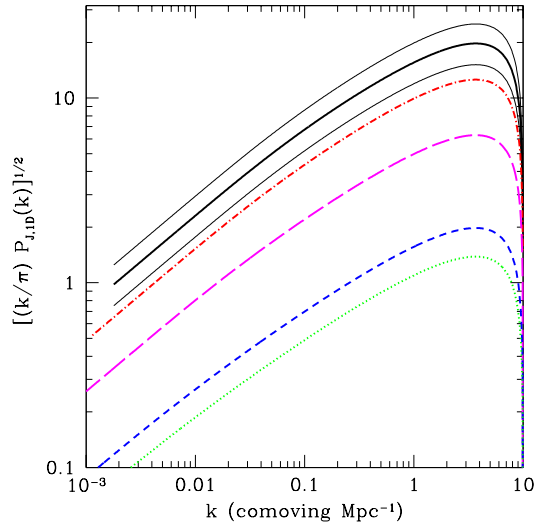}
\caption{Fractional rms scatter in the radiation field along one-dimensional skewers, as a function of smoothing scale (or wavenumber $k$).  All thick curves assume $z=2.5$ and take $r_0=35,\,55,\,110,\,350,$ and 500 Mpc, from top to bottom (line styles are identical to Fig.~\ref{fig:corr}); the thin solid curves show the plausible range of uncertainty from $\Phi(L)$.}
\label{fig:pk_1d}
\end{figure}

Figure~\ref{fig:pk_1d} shows the one-dimensional power spectrum of the ionizing background, again presented in non-dimensional form such that the amplitude is the rms scatter expected when data are smoothed on a given physical scale.  The curves are identical to those in Figure~\ref{fig:pk}.  The exponential suppression at $k= 10 \Mpcinv$ reflects the assumed Jeans scale.  Because of aliasing, the one-dimensional variance is always dominated by the smallest scales, and there is only a slight break at $r_0$ in $P_{\rm J,1D}(k)$; it will unfortunately be very difficult to extract this parameter from one-dimensional \lya forest data.

However, the amplitude of $P_{J,\rm 1D}$ does depend strongly on $r_0$ and offers another route to estimate it, albeit with relatively large uncertainties.  Unfortunately, the plausible range of $\Phi(L)$ (illustrated by the thin solid curves), introduces about a factor of two uncertainty in the overall amplitude and will make it difficult to measure $r_0$ robustly.

We also find that the fluctuation amplitudes are quite large:  for $r_0=35 \Mpc$, the fractional variations are of order unity even across hundreds of Mpc, and an order of magnitude at scales $\sim 20 \Mpc$.  In contrast, the variations in the three-dimensional field only reach this level at $\sim 6 \Mpc$.  Even when $r_0=500 \Mpc$, order unity fluctuations appear at $\sim 10 \Mpc$.

\section{Discussion}
\label{disc}

We have shown that the scarcity of quasars induces strong correlations in the radiation background, even if those sources are themselves randomly distributed.  This is especially important for \ion{He}{2}-ionizing photons, whose attenuation lengths are $\la 50 \Mpc$ after \ion{He}{2} reionization, and so have expected correlation lengths $\sim 25 \Mpc$.  The variations are especially large when viewed along one-dimensional skewers (as provided by the \lya forest):  in those circumstances aliasing induces order-unity fluctuations on scales $\ga 100 \Mpc$, and order-of-magnitude fluctuations on scales $\la 20 \Mpc$. 

\subsection{Comparison to Observations and Simulations}
\label{comp}

The best way to observe these correlations is by comparing the \ion{He}{2} and \ion{H}{1} \lya forests.  The relative abundance of these species depends primarily on the relative radiation backgrounds above their ionization thresholds, so the hardness parameter $\eta = N_{\rm HeII}/N_{\rm HI}$ provides a measure of the local strength of the \ion{He}{2} ionization rate \citep{miralda93, shull04, zheng04, fechner06, fechner07}.  In the limit of a uniform $\Gamma_{\rm HI}$, and if the line comparison can be made cleanly, $\eta \propto 1/\Gamma_{\rm HeII}$.\footnote{We should note that it is not trivial to compare lines in this way, especially if they are thermally broadened \citep{fechner07-therm}, and in some cases the ``pixel optical depth method" may be superior.  Fortunately, the two methods appear to agree in most cases \citep{fechner07}.  Direct comparisons to numerical simulations is probably the safest route, but it has not yet been possible (largely because of the difficulty of properly simulating \ion{He}{2} reionization; \citealt{paschos08, mcquinn08}).}  In that case, we can compute $\xi_J$ directly from the observations because $\xi_{1/\eta} = \xi_J$ \citep{zuo92b}.

Although this correlation function has not yet been measured precisely, the data clearly show large $\eta$ variations on rather small scales.  For example, \citet{shull04} detected substantial $\eta$ fluctuations over $\sim 2 \Mpc$ segments toward HE 2347--4342.  But their quantitative level remains unclear. For example, \citet{fechner07} showed that the highest signal-to-noise portions of that spectrum, and the entire line of sight to HS 1700+6416 (also high signal-to-noise), are dominated by smoother variations, with $\sim 33\%$ of the IGM varying on scales $\la 6 \Mpc$ and $\ga 50\%$ varying only on scales $\ga 14 \Mpc$.  

Contrary to previous claims, these qualitative measurements are consistent with the large variations on moderate and small scales expected from a simple model of discrete quasar sources, even though the attenuation length is much larger.  In particular, we emphasize that aliasing amplifies the apparent fluctuations along \lya forest skewers, so other processes -- such as complex radiative transfer -- may not be required to reproduce the observations (although they may still be important; \citealt{maselli05, tittley07}).  

Quasar-induced fluctuations have been directly detected through the ``transverse proximity effect" in $\eta$ along two separate lines of sight \citep{jakobsen03, worseck06, worseck07}:  the radiation background hardens near several quasars close to the line of sight.  However, even within the regions of low $\eta$, there are still large fluctuations, especially at the lower redshifts (where $r_0$ is larger), and there are other regions with low $\eta$ but no sources within $\sim 30 \Mpc$.  Whether these differences are due to aliasing, anisotropic emission, or radiative transfer remains to be seen.

Although detailed numerical simulations have not yet addressed this question, support for our conclusion was provided by \citet{bolton06}, who found strong small-scale variations in simulated data using a model of discrete quasar sources without appealing to true radiative transfer (see their Fig. 5).  The next step is to compute the power spectrum of the data in order to quantitatively compare to models like ours.  Whether that is possible with the sparse and noisy data currently available remains to be seen, but many more lines of sight have recently become available \citep{zheng08, syphers08}, so these tests should sharpen in the near future.

\subsection{Caveats to the Model}
\label{caveat}

However, at the moment our model itself is also too crude to make detailed comparisons with the data, and improvements are clearly needed on the theoretical side.  In particular, several effects will conspire to \emph{decrease} the observed fluctuations relative to our calculations.  First, we have included the entire intensity profile around each source, including the (divergent) $r \rightarrow 0$ limit.  Of course, structure within the quasars host halo will actually modify the small-scale profile (and it is not measured by the \lya forest anyway), decreasing some of the high-$k$ power.  

More importantly, we have ignored the limited dynamic range available in measurements of $\eta$:  in reality, we cannot measure either extremely large values of $\eta$ (because of saturated absorption) or extremely small values (where the transmission is essentially unity).  The latter limit is especially important, because it will damp the observed power within the highly-transparent central proximity zone of each source.  \citet{zuo92b} considered some of the complexities of the nonlinear transformation, but this effect is best studied with mock spectra as in \citet{bolton06}.

Third, we have only considered photons with a single, well-specified attenuation length (and chosen $r_0$ for photons just above the \ion{He}{2} ionization edge).  However, quasars have hard spectra (typically with $L_\nu \propto \nu^{-1.6}$ in the far-UV; \citealt{telfer02}) and so produce a substantial number of high-energy photons.  Because such photons have longer mean free paths, they damp the correlations \citep{furl08-hefluc}.  However, the photoionization cross section falls steeply with frequency ($\sigma_\nu \propto \nu^{-3}$), so this should not qualitatively change our conclusions.

Another concern is our assumption that the entire \ion{He}{2}-ionizing background comes from quasars.  While galaxies are unlikely to be significant (but see \citealt{furl08-helium}), a diffuse background must develop from the recombination radiation that follows IGM ionizations.  The importance of this background depends on the geometry of the absorbers:  provided that most are optically thick regions, much of the recombination radiation will be directed deeper into the (mostly neutral) system and thus not influence the IGM \citep{miralda03}.  However, in the opposite case-B limit, $\sim 40\%$ of the photons are re-emitted.  This could therefore decrease the amplitude of $\xi_J$ by a factor of order unity.

One might also wonder whether the assumption of a uniform $\Gamma_{\rm HI}$ is acceptable for these purposes:  inside the proximity zone of each quasar, both $\Gamma_{\rm HeII}$ and $\Gamma_{\rm HI}$ increase, such that $\eta$ may remain constant.  However,  we have shown that, for the large attenuation lengths relevant to \ion{H}{1}-ionizing photons, the correlations induced by discrete quasars only manifest themselves on $\la 1 \Mpc$ scales, and this will be reduced still farther if galaxies dominate the \ion{H}{1}-ionizing background (as now seems likely; e.g., \citealt{faucher08-ionbkgd}).  This is over an order of magnitude smaller than the correlation length for \ion{He}{2}-ionizing photons.  In other words, because so many more sources contribute to the local \ion{H}{1}-ionizing field, the \ion{H}{1} proximity zone is much smaller than that for \ion{He}{2}, so it cannot compensate for the fluctuations in the higher-energy background.

On the other hand, we have ignored the deterministic bias of quasar hosts and included the strong variations seen in far-UV quasar spectra only crudely \citep{telfer02, scott04}.  Both of these effects will amplify the small-scale correlations, although the degree remains uncertain.

The large fluctuations we have found may seem surprising in comparison to previous studies, such as \citet{meiksin03}.  They considered a quasar-dominated ionizing background at high redshifts ($z \ga 4$), where the fluctuations in $J$ are also large.  But they found that their effects on the \ion{H}{1} \lya forest transmission statistics (such as the flux probability distribution function and its power spectrum) remain modest.  Several aspects of those statistics probably account for the difference with our conclusion.  First, the \lya optical depth, on its own, is most sensitive to IGM density fluctuations, which may mask variations in the radiation background (e.g., \citealt{worseck06}).  Second, the complex line structure of the forest (peculiar velocities, line broadening, etc.) may conceal some of the $\Gamma$ fluctuations.  Finally, the $e^{-\tau}$ suppression decreases the available dynamic range in the forest, as described above.  Direct comparison of the \ion{He}{2} and \ion{H}{1} forests provides a cleaner tool to study these fluctuations principally because it removes the degeneracy with density (which affects both lines the same way, at least in principle).  \citet{meiksin03} also assumed a significantly steeper luminosity function ($\beta=-3.41$) than ours ($\beta \approx -2.9$), which reduces the fluctuations by a factor $\sim 2$.

\subsection{Future Directions}
\label{future}

In addition to the improvements outlined above, spatial correlations in the high-energy radiation field can be used to address two other sets of questions.  First, we have only calculated $\xi_J$ after \ion{He}{2} reionization has ended.  During reionization, the ionizing background fluctuates even more strongly \citep{furl08-hefluc}, with fewer discrete sources illuminating any point in the IGM.  Thus the correlations will be even stronger -- and, indeed, in many ionized regions there will only be a single active quasar, which could allow us to trace out its region of influence in detail.

Second, to the extent that these correlations reflect the proximity zones of individual bright quasars, we may be able to pick out regions with large $\Gamma_{\rm HeII}$ from the observed spectra and associate them with known quasar (or at least galaxy) locations.  In the best case, we could trace the intensity profile around each nearby source and so test for the ``transverse proximity effect;" such tests have already shown hardening in the local radiation field near quasars and placed bounds on the lifetimes of UV-bright quasars \citep{worseck06, worseck07}.  In the crudest sense, such regions would suggest target regions for future quasar searches (as in \citealt{jakobsen03}).  These studies require high signal-to-noise measurements of the \lya forest (in both \ion{He}{2} and \ion{H}{1}) and also detailed maps of their environments, ideally spanning a couple of attenuation lengths (or $\sim 100 \Mpc$, several times farther than \citealt{worseck06, worseck07}).  This exercise would elucidate, in detail, not only the attenuation of high-energy photons throughout the IGM but also properties of quasar host galaxies, the lifetime and light curves of quasar sources (by tracing the intensity profiles of each source), and the geometry of the quasar emission (i.e., whether it is isotropic and whether it evolves over the lifetime of the quasar).  

\acknowledgments

I thank A.~Lidz for comments that greatly improved this manuscript.  This research was partially supported by the NSF through grant AST-0607470 and by the David and Lucile Packard Foundation.


\end{document}